\documentclass[12pt,preprint,showpacs,showkeys,amssymb,amsmath,aps]{revtex4}
\usepackage{epsfig}
\usepackage{bm}
\usepackage{docs}

\begin{document}
\title{From Local Regge Calculus towards Spin Foam Formalism?}

\author{Gabriele Gionti S.J.}
\affiliation{Vatican Observatory Research Group,
Steward Observatory, 933 North Cherry Avenue, The University of Arizona, Tucson AZ 85721 U.S.A.}
\affiliation{Specola Vaticana, V-00120 Citt\'a Del Vaticano, Vatican City State}
\email{ggionti@as.arizona.edu}

\begin{abstract}
We introduce the basic elements of $SO(n)$-local theory of Regge Calculus. A first order
formalism, in the sense of Palatini, is defined on the metric-dual Voronoi complex
of a simplicial complex. The Quantum Measure exhibits an expansion, in four dimensions, 
in characters of irreducible representation of $SO(4)$
which has close resemblance and differences as well with the Spin Foam Formalism. The 
coupling with fermionic matter is easily introduced which could have consequences for the 
Spin Foam Formalism and Loop Quantum Gravity.   
\end{abstract}


\maketitle

\section{Introduction}
Regge Calculus \cite{regge} (see also \cite{ruth} for a recent and updated summary
of Regge Calculus and its alternative approaches)  provides a discretized version of General
Relativity. It is, mainly, based on the idea
of substituting a  Piecewise-Linear (PL)-manifold for differential manifold. In
particular, these are simplicial manifolds built by gluing together two distinct
$n$-dimensional flat simplexes by one and only one ($n-1$)-dimensional
face. The final product of this construction is called a simplicial complex,
which owns the manifold structure if it has been
made in such a way that each point of the simplicial complex  has  a
neighbourhood homeomorphic to ${\Re}^{n}$ \cite{seifert}. The simplicial manifolds,
 we consider, are orientable. On a simplicial manifold we can define a metric
structure. The metric
is Euclidean almost everywhere except near the ($n-2$)-simplexes called hinges
$h$. Near the hinges the metric is equivalent to a {\it cone}-metric,
that is the metric {\bf g} of $\Re^{n-2}
\times$ the two dimensional cone \cite{thurston}
\cite{regge}.
 A two-dimensional cone is formed by cutting away a
circular sector of angle $\theta$ and gluing together the
two free edges. Its apex is called cone point of angle $\theta$
and of curvature $2\pi - \theta$.
On each ($n-2$)-dimensional
simplex $h$, called an {\it hinge}, a deficit angle $K(h)$, which
is the cone curvature,  is defined

\begin{equation}
K(h)= 2\pi - \sum_{\sigma^{i}_{h} \supset h}\theta(\sigma^{i}_{h},h)
\label{apex}
\end{equation}

\noindent where $\sigma^{i}_{h}$, $i=1,...,p$ is one of the $p$
$n$-dimensional simplices incident on $h$, and
$\theta(\sigma_{h}^{i},h)$ its dihedral angles on $h$.
The dihedral angle of a $n$-simplex on the hinge
is the angle between the ($n-1$)-dimentional faces that have the hinge in common.
$K(h)$ and the volume of the hinge $V(h)$ can be expressed \cite{hamber}
as functions of the squared length of the one-dimensional simplices (edges) of
the complex.
The Einstein-Regge action is, in analogy to the continuum case,
a functional over
the simplicial manifolds and depends on the incidence matrix of the simplicial
complex \cite{frohlich} and on the (squared) lengths of the edges. Usually, in
Regge calculus, the incidence matrix is fixed, so that the action can be written as

\begin{equation}
S_{R}=\sum_{h}K(h)V(h)
\label{reggeact}\;\;\;\; .
\end{equation}

\noindent The corresponding
Einstein equations \cite{regge} are derived
by requiring that the action is stationary
under the variation of the length of the edges.

\noindent The original aim of this theory was to give approximate solutions of
the Einstein equations in the case in which the topology is known. The theory is, as
stressed by Regge himself, completely coordinate independent.

\begin{figure}

\epsfxsize=8.0truecm
\epsfysize=8.0truecm
\centerline{\hbox{\epsffile{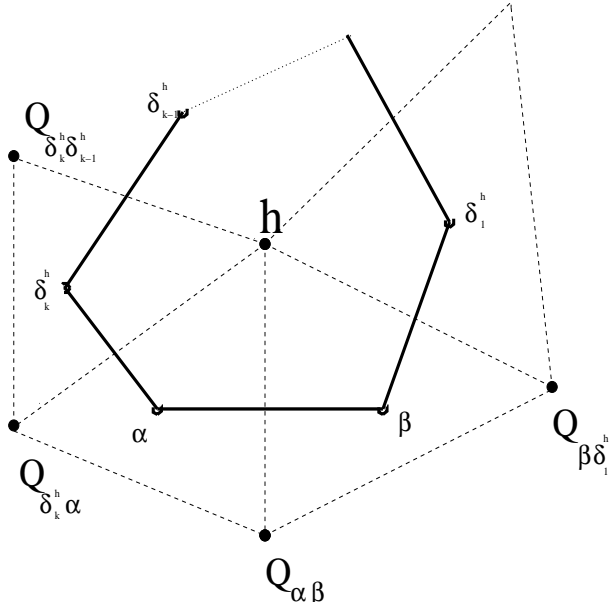}}}

\caption{\label{giorno1} Dual Voronoi Plaquette}

\end{figure}

\noindent Now we summarize and update  some definitions and results of \cite{Ale} that
will be useful for our future discussions.
Consider an hinge $h$ and suppose that
it is shared by $k+2$ $n$-simplices (see FIG.1)
$\{\alpha,\beta,\delta^{h}_{1},...,\delta^{h}_{k}\}$. 

\noindent Given a $n$-dimensional simplicial complex, there exists a general procedure for defining the
dual metric complex, called the Voronoi dual \cite{Ale}
\cite{senechal}. The $n$-dimensional Voronoi polyhedron dual to a vertex $P$
of the simplicial complex is the set of the points of the simplicial complex closer to $P$
than to any other vertex, using the standard flat metric in the simplex. It turns out that the $k$-simplex
of the simplicial complex is dual to a $(n-k)$-polyhedron in the dual Voronoi complex , and the $k$-dimensional linear
space identified by the $k$ simplex is orthogonal to the $(n-k)$-dimensional space spanned by the corresponding
polyhedron. In particular the point dual to the $n$-simplex $S$ is the centre of a $(d-1)$-dimensional sphere
passing through all vertices of $S$ (look at FIG.1).

\noindent It was pointed out in reference \cite{Ale} that a theory of Regge Calculus locally invariant
under the Poincar\'e group can be formulated
by choosing in every simplex an orthonormal reference frame.
In this
way every $n$-dimensional simplex, considered as a piece of ${\Re}^{n}$, can be
seen as local inertial reference frame, being flat, and an n-bein base can be
chosen in it. As in the continuum theory, we need a connection matrix between
the reference frames of two simplexes that share a common ($n-1$)-dimensional simplex.
Fixing \cite{Ale} \cite{thesis} \cite{Gab} the origin of the reference frames on the dual Voronoi vertices,
the orthogonal matrix $\Lambda^{a}_{b}(\alpha,\beta)$, which allows passing from the reference frame of $\alpha$
to the $\beta$'s, happens to be the Levi-Civita \cite{frohlich}, Torsion-free, connection matrix which is a 
function of the vertex coordinates
of the common face $\alpha \beta$ both in the reference frame of $\alpha$ and $\beta$.

\noindent The hinge $h$ is in one-to-one correspondence
with its dual two-dimensional Voronoi
plaquette that we still label by $h$.
Now consider
the following holonomy matrix and plaquette variable

\begin{equation}
W_{\alpha}(h)=\Lambda(\alpha,\beta)
\Lambda(\beta,\delta^{h}_{1})...\Lambda(\delta^{h}_{k},\alpha)\;\;\;\; .
\label{ruota}
\end{equation}

\noindent As it has been shown \cite{hamber} \cite{thesis}, (\ref{ruota}) is a rotation
in the two-dimensional plane orthogonal to the ($n-2$)-dimensional hyperplane
spanned by the hinge $h$. The rotation angle is the deficit angle (\ref{apex}).

\noindent We consider $n-2$
linear independent edge-vectors of the hinge $h$ defined in the following way

\begin{equation}
E_{1}^{a}(\alpha)\equiv y^{a}_{1}(\alpha)-y^{a}_{n-1}(\alpha),...,
E_{n-2}^{a}(\alpha)\equiv y^{a}_{n-2}(\alpha)-y^{a}_{n-1}(\alpha),
\label{vecto}
\end{equation}

\noindent and the antisymmetric tensor related to the oriented volume of $h$

\begin{equation}
{\mathcal V}^{(h)}{\;}^{ab}(\alpha)\equiv \frac{1}{(n-2)!}
\epsilon^{ab}_{c_{1}...c_{n-2}}E_{1}^{c_1}(\alpha)...E_{n-2}^{c_{n-2}}(\alpha)\;\;\;\; .
\label{orio}
\end{equation}

\noindent At this point it seems natural to propose the following gravitational action \cite{Ale}

\begin{equation}
I=-{1\over 2}\sum_{h}\Bigl( W_{\alpha}^{a_{1}a_{2}}{\;}^{(h)}
{\mathcal V}^{(h)}{\;}_{a_{1}a_{2}}(\alpha)\Bigr)
\label{haction}\;\;\;\;.
\end{equation}

\noindent Let $V(h)$ be the oriented volume of the hinge $h$.
The action (\ref{haction}) is equal \cite{Ale} to

\begin{equation}
I= \sum_{h}sin\;K(h)\;V(h)\;\;\;\; ,
\label{calc}
\end{equation}

\noindent which for small deficit angles $K(h)$ reduces to the Regge action (\ref{reggeact}).

\noindent The presence in the action of $sinK(h)$ instead of $K(h)$ is only a
lattice artifact \cite{Ale}. The useful regime of this theory is $sinK(h) \approx K(h)$. In fact
Fr\"ohlich \cite{frohlich} has conjectured, as claimed also in \cite{Ale}, that this action converges to the Einstein-Hilbert action,
and J.Cheeger, W. Muller, R.Schrader in \cite{Cheeger} have proved that the Einstein-Regge action (\ref{reggeact})
converges to the Einstein-Hilbert action. They proved that the convergence of the action is in {\it sense of measure}.
This applies when, roughly speaking,
the number of the hinges of the simplicial manifold increases along with the incident number of the simplices at each hinge.
Then the triangulations will be finer and finer and the difference, in modulus, between the Einstein-Hilbert action on a manifold and
the Regge-Einstein one, on the triangulations of the same manifold, become smaller and smaller.

\section{First-order formalism}

Let's consider the ($n-1$)-dimensional face $f_{\alpha\beta}\equiv \{P_1,...,Q_{\alpha\beta}\}$
between the simplices $\alpha$ and $\beta$. The vertex coordinates of $f_{\alpha\beta}$
are $x_{i}^{a}(\alpha), a=1,...,n$. The following vector \cite{Ale}

\begin{eqnarray}
{b_{\alpha\beta}}{\;}_{a}(\alpha)&=&
\epsilon_{ab_{1}...b_{n-1}}(x_{1}}(\alpha)-x_{n}(\alpha))^{b_{1}}...
(x_{n-1}(\alpha)-x_{n}(\alpha))^{b_{n-1}\nonumber\\
&=&
\epsilon_{ab_{1}...b_{n-1}}E^{b_{1}}_{1}...E^{b_{n-1}}_{n-1}
\label{1tetra}
\end{eqnarray}

\noindent is normal to the face $f_{\alpha\beta}$, and it is assumed
to point outward from the interior of the simplex $\alpha$.
The analogous vector ${b_{\beta\alpha}}{\;}_{a}(\beta)$ in the reference frame of $\beta$ is related
to the previous one by

\begin{equation}
{b_{\alpha\beta}}{\;}^{a}(\alpha)=
 \Lambda^{a}_{b}(\alpha,\beta){b_{\beta\alpha}}{\;}^{b}(\beta)\;\;\;\; .
\label{relatto}
\end{equation}

\noindent  It can be easily proved that (\ref{orio}) can be written as a bivector

\begin{equation}
{\mathcal{V}}^{(h)}{\;}^{c_1c_2}(\alpha)\equiv
{1\over n!(n-2)!V(\alpha)}\Bigl({b_{\alpha\beta}}^{c_1}
(\alpha){b_{\alpha\delta^{h}_{k}}}^{c_2}(\alpha)
-{b_{\alpha\beta}}^{c_2}(\alpha){b_{\alpha,\delta^{h}_{k}}}^{c_1}(\alpha)\Bigr),
\label{arrota}
\end{equation}

\noindent where $V(\alpha)$ is the oriented volume of the simplex $\alpha$.

 We can consider the action as written on the dual Voronoi-complex of the
original simplicial complex. The dual of the hinge $h$ is a two dimensional
plaquette whose vertices will be
$\left\{\alpha,\beta,\delta^{h}_{1},...,\delta^{h}_{k}\right\}$. These vertices are the dual
of the $n$-simplices incident on the hinge $h$.

\noindent If, briefly, we indicate $\Lambda(\alpha,\beta)$ as $\Lambda_{\alpha\beta}$, and so on, the holonomy
matrix (\ref{ruota}) around the plaquette is

\begin{equation}
U^{h}_{\alpha \alpha}\equiv
\Lambda_{\alpha\beta}\Lambda_{\beta\delta^{h}_{1}}
...\Lambda_{\delta^{h}_{k}\alpha}\;\;\;\;,
\label{holonomy}
\end{equation}

\noindent However this is still a second order formalism of discrete General Relativity .
The connection matrices $\Lambda_{\alpha\beta}$ and the $b_{\alpha\beta}$
are both functions of the coordinates of the edges of the simplicial complex.
Now we introduce a first
order formalism in which $\Lambda_{\alpha\beta}$ and
$b_{\alpha\beta}(\alpha)$ are independent variables. Then equation (\ref{relatto}) becomes a constraint

\noindent On the $b_{\alpha\beta}(\alpha)$ there is a further constraint
 since  the $n+1$ normals to the ($n-1$)-dimensional faces
 of a $n$-simplex $\alpha$ are linearly dependent,

\begin{equation}
\sum_{\beta=1}^{n+1}b_{\alpha\beta}(\alpha)=0
\label{chiusura}\;\;\;\; .
\end{equation}

 \noindent There is a technical problem we would like pointing out.
The gravitational action
in the form (\ref{haction}) could be
dependent on the starting simplex .
 So we need to define the following
antisymmetric tensor on the plaquette $h$

\begin{eqnarray}
W^{(h)}_{c_1c_2}(\alpha) &\equiv &
{1\over k_{h}+2} \Big({\mathcal{V}}^{(h)}(\alpha)+
\Lambda_{\alpha\beta}{\mathcal{V}}^{(h)}(\beta)\Lambda_{\beta\alpha}+... \nonumber\\
&+&\Lambda_{\alpha\beta}...
\Lambda_{\delta^{h}_{k-1}\delta^{h}_{k}}
{\mathcal{V}}^{(h)}(\delta^{h}_{k})\Lambda_{\delta^{h}_{k}\delta^{h}_{k-1}}...
\Lambda_{\beta\alpha} \Big)_{c_1c_2}
\label{media}\;\;\;\;.
\end{eqnarray}

\noindent The action then is

\begin{equation}
S\equiv -{1\over 2}\sum_{h}{\rm Tr}
\left(U^{h}_{\alpha\alpha}W^{h}(\alpha)\right)
\label{caction}
\end{equation}

\noindent The action (\ref{caction}) coincides with the action (\ref{haction}) in the second-order formalism, and,
since (in matrix notation)

\begin{equation}
W^{(h)}(\alpha)=\Lambda_{\alpha\beta}...\Lambda_{\delta_{i-1}\delta_{i}}
W^{(h)}(\delta_{i})\Lambda_{\delta_{i}\delta_{i-1}}...\Lambda_{\beta\alpha}
\label{1lastessa}\;\;\;\;,
\end{equation}

\noindent equation (\ref{caction}) is  independent of the starting simplex.

\noindent Moreover the action (\ref{caction}) is invariant under the following set of
transformations

\begin{eqnarray}
\Lambda_{\alpha\beta}&\mapsto&
{\Lambda}'_{\alpha\beta}=
O(\alpha)\Lambda_{\alpha\beta}O^{-1}(\beta)\nonumber\\
b_{\alpha\beta}(\alpha)&\mapsto&
b'_{\alpha\beta}(\alpha)=O(\alpha)b_{\alpha\beta}(\alpha)
\label{gauge}
\end{eqnarray}

\noindent where $O(\alpha)$ and $O(\beta)$ are two elements of $SO(n)$. This can be interpreted as the diffeomorphism
invariance of this discrete theory of Gravity.

\section{From Quantum Measure to Spin Foam}

In this section we shall mainly focus on the four
dimensional case. The action (\ref{caction}) is invariant under the group $SO(4)$ which
is the gauge group. It has been proved \cite{thesis} \cite{Gab} that the discrete equations of motion for this theory 
can be derived and appear very similar to General Relativity equations in the covariant tetrad formalism \cite{myself},
and in the limit for {\it small deficit angles} the {\it Levi-Civita-Regge} connection (the second order formalism which,
in this case, is the Regge Calculus itself) is, locally, the unique solution. 
This is, clearly, an analogy with the Palatini continuum theory
where, in the Torsion-free case, the Levi-Civita connection is solution of one part 
of the first order field equations.

\noindent We will use the following notation:

\begin{equation}
\mu(b_{\alpha\beta}(\alpha))\equiv
db^{1}_{\alpha\beta}(\alpha)...db^{n}_{\alpha\beta}(\alpha)
\label{bmes}\;\;\;\;,
\end{equation}

\noindent and let

\begin{equation}
\mu(\Lambda_{\alpha\beta})
\label{Haar}
\end{equation}

\noindent be  the Haar measure on $SO(4)$. The partition function for this theory is:

\begin{equation}
Z=\int e^{{1\over 2}\sum_{h}Tr\left(U^{h}_{\alpha\alpha}W^{h}(\alpha)\right)}
\prod_{\alpha}\delta(\sum_{\beta=1}^{n+1}b_{\alpha\beta}(\alpha))
\prod_{\alpha\beta}
\delta\left(b_{\alpha\beta}(\alpha)-\Lambda_{\alpha\beta}b_{\beta\alpha}(\beta)\right)
\mu(\Lambda_{\alpha\beta})\mu(b_{\alpha\beta})\;\;\;\;.
\label{part}
\end{equation}

\noindent And this path-integral is invariant \cite{Gab} under local $SO(4)$ transformations (\ref{gauge}).

From Group Theory \cite{vilenkin} it is well known that a function $f$ on a compact group $G$ is called central
if $\forall g, h \in G, f(h^{-1}gh)=f(g)$. Thanks to the gauge invariance (\ref{gauge}) previously mentioned  $e^{{1\over
2}Tr\left(U^{h}_{\alpha\alpha}W^{h}(\alpha)\right)}$ is a central function on $SO(4)$. It follows
\cite{vilenkin} we can expand it in terms of the characters $\chi_{J}\left(U^{h}\right)$ of the irreducible
representations $J$ of the group $SO(4)$. We can omit the index $\alpha\alpha$ in the holonomy
matrix $U^{h}$ since it does not depend from the starting simplex $\alpha$.

\begin{equation}
e^{{1\over 2}Tr\left(U^{h}_{\alpha\alpha}W^{h}(\alpha)\right)}=\sum_{J}
c_{J}^{h}\Big(b_{\alpha\beta},b_{\beta\alpha},...,
b_{\delta^{h}_{k}\alpha},b_{\alpha\delta^{h}_{k}}\Big)\chi_{J}\left(U^{h}\right)\;\;\;\;,
\label{svilup}
\end{equation}

\noindent where the coefficients $c_{J}^{h}\Big(b_{\alpha\beta},b_{\beta\alpha},...,
b_{\delta^{h}_{k}\alpha},b_{\alpha\delta^{h}_{k}}\Big)$ are defined by \cite{vilenkin}

\begin{equation}
c_{J}^{h}\Big(b_{\alpha\beta},b_{\beta\alpha},...,b_{\delta^{h}_{k}\alpha},b_{\alpha\delta^{h}_{k}}\Big)\equiv\int_{SO(4)}\mu\big(U^{h}\big)e^{{1\over
2}Tr\left(U^{h}_{\alpha\alpha}W^{h}(\alpha)\right)}
{\overline{\chi_{J}\left(U^{h}\right)}}\;\;\;\;,
\label{coeffio}
\end{equation}

\noindent $\mu\big(U^{h}\big)$ being the $SO(4)$ Haar measure 
on the holonomy matrices $\big(U^{h}\big)$ and 
$\overline{\chi_{J}\left(U^{h}\right)}$ the complex conjugate of 
$\chi_{J}\left(U^{h}\right)$. 

\noindent These new considerations imply we can write the previous Quantum Measure as

\begin{eqnarray}
Z=\int \prod_{h}\sum_{J}c_{J}^{h}\Big(b_{\alpha\beta},b_{\beta\alpha},...,
b_{\delta^{h}_{k}\alpha},b_{\alpha\delta^{h}_{k}}\Big)\chi_{J}\left(U^{h}\right)
\prod_{\alpha}\delta\big(\sum_{\beta=1}^{n+1}b_{\alpha\beta}(\alpha)\big)\nonumber\\
\prod_{\alpha\beta}
\delta\big(b_{\alpha\beta}(\alpha)-\Lambda_{\alpha\beta}b_{\beta\alpha}(\beta)\big)
\mu(\Lambda_{\alpha\beta})\mu(b_{\alpha\beta})\;\;\;\;,
\label{color}
\end{eqnarray}

\noindent and this last equation seems suggesting that we can label the interior of 
each plaquette by an irreducible representation $J$ of $SO(4)$. This is very similar to the Spin Foam Formalism where the 
interior of each oriented $2$-polyhedron is labeled, or colored, by an irreducible representation of 
$SO(4)$ \cite{rovelli}\cite{Oritith}.Although this is an early first stage result some sensible differences appear
respect to Spin Foam Formalism. 
The main one, we would like to highlight, is that Barret and Crane \cite{barrettcrane} proved that the geometry
of each four-simplex can be completely characterized by ten bivector two-forms   

\begin{equation}
B_{IJ}\equiv E_{I}\wedge E_{J}
\label{bivoco}
\end{equation}

\noindent where $E_{I}$ and $E_{J}$ are two independent edge vectors which 
belong to one of ten triangles of a four-dimensional 
simplex. In order to label 
each oriented $2$-polyhedron by an irreducible representation of $SO(4)$ they quantize the 
bivector two-form. Then they associate, in four dimensions, an irreducible representation $J$ to the interior of each 
triangle. Some technical problems arise in implementing this quantization\cite{reisenberger}. Mainly in the simplicial $BF$ theory for Spin Foam Formalism, 
which is the proposal of 
Barret-Crane model\cite{barrettcrane}, it is needed to consider a generic two form $B_{IJ}$ and impose, at quantum level, that equation 
(\ref{bivoco}) be a quantum constraint.        

\noindent Although we have also a bivector two-form $W^{h}$ (\ref{media}) in the action (\ref{caction}), which has a close resemblance with $BF$
and Plebanski's action for General Relativity, it is very clearly defined on the $2$- 
Voronoi-plaquette instead of the triangle dual to the plaquette. Moreover we do not
need to quantize this bivector two-form because, as we have seen, the expansion (\ref{svilup}) 
in the path-integral tells us how to label the plaquette by an irreducible representation of $SO(4)$.

Anyway, till now, we have not found that the partition function (\ref{color}) can be written as claimed 
in Spin Foam Formalism \cite{rovelli}:

\begin{equation}   
Z =\sum_{\Gamma}w(\Gamma)\sum_{J_{f},i_{e}}
\prod_{f}A_{f}(j_{f})
\prod_{e}A_{e}(j_{f},i_{e})
\prod_{v}A_{v}(j_{f},i_{e})\;\;\;\; .
\label{foam}
\end{equation}

\noindent where $\Gamma$ is an oriented $2$-complex, $w(\Gamma)$ is a weight associated to each $2$-complex $\Gamma$, $J_{f}$ is 
an $SO(4)$-irreducible representation associated to each $2$-face $f$, $i_{e}$ an intertwiner which, at each edge $e$, maps the $SO(4)$-irreducible 
representations associated with the incoming $2$-faces at the edge $e$  
to the outcoming $SO(4)$-irreducible representations. $A_{f}(J_{f})$ is a coefficient 
associated wih each face $f$ and so $A_{e}(j_{f},i_{e})$ and $A_{v}(j_{f},i_{e})$ 
with each edge $e$ and vertex $v$ respectively. One of the reason of it, we have found, 
is that the constraints (\ref{relatto}) and (\ref{chiusura}),
the delta function of the partition function (\ref{color}), make the integrals highly 
non-local objects with some similarities with the measure of
Quantum Regge Calculus \cite{menotti1}. It happens that the $b^{i}$'s, which are not adjacent and ``very far'', 
can be correlated by implementing the constraints.This 
seems to prevent writing the partition functions (\ref{color})  as a product of three factor as in (\ref{foam}). 
In BC-model, as well as other models for Spin Foam 
Formalism \cite{rovelli}, the four-dimesional simplices, or equivalently, in the dual map, the dual vertices,  
are, in general, separeted one form the others and there are no constraints to be implemented 
so the partition function results to be  factorized as in (\ref{foam}).

\section{Coupling with fermionic  matter}

\noindent  The coupling with fermionic matter, in the continuum case 
on Riemannian manifolds with torsion-free
connection, is given
by the lagrangian density

\begin{equation}
{\mathcal L}\equiv \frac{i}{2}\Big({\overline {\psi}}e^{\mu}_{a}\gamma^{a}{\nabla}_{\mu}\psi
- e^{\mu}_{a}\big(\overline{{\nabla}_{\mu}{\psi}}\big)\gamma^{a}\psi\Big)
-m\psi {\overline \psi}
\label{lagra}\;\;\;\;.
\end{equation}

\noindent The  $\gamma^{a},\; a=1,...,n$ are the Dirac-matrices satisfying the Clifford
algebra

\begin{equation}
\gamma^{a}\gamma^{b}+\gamma^{b}\gamma^{a}=2\delta^{a\;b}
\label{gamma}\;\;\;\;,
\end{equation}

\noindent whereas $\psi$ is  the n-dimensional Dirac spinor field
(${\bar \psi}\equiv {\psi}^{\dagger}{\gamma}^{1}$),${\nabla}_{\mu}$ the
covariant derivative, and $e_{a}^{\mu}$ the n-beins
on the tangent space of the Riemannian manifold $(M,g)$.

\noindent Now we have all the ingredients to define the coupling of gravity with
fermionic matter on the lattice in analogy with the continuum case\cite{ren}.
Let $2\nu=n$ or $n=2\nu+1$ (depending on whether
$n$ is even or odd) and consider the $2^{\nu}$-dimensional representation
of the two-fold covering group of $SO(n)$\cite{frohlich}. So, instead of considering the
connection matrices $\Lambda_{\alpha\beta}$, we will deal with the
$2^{\nu}\times 2^{\nu}$ connection matrices $D_{\alpha\beta}$ such that

\begin{equation}
D_{\alpha\beta}{\gamma}^{a}D_{\alpha\beta}^{-1}
=(\Lambda_{\alpha\beta})^{a}_{b}{\gamma}^{b}
\label{2-fold}\;\;\;\;.
\end{equation}

\noindent Given $D_{\alpha\beta}$ we can determine
$\Lambda_{\alpha\beta}$. Furthermore if we know $\Lambda_{\alpha\beta}$, we can determine
$D_{\alpha\beta}$ up to a sign. In particular, from
(\ref{2-fold}), we can write $\Lambda_{\alpha\beta}$ as \cite{frohlich}

\begin{equation}
(\Lambda_{\alpha\beta})_{ab}={1\over
2^{\nu}}Tr(\gamma_{a}D_{\alpha\beta}\gamma_{b}D_{\alpha\beta}^{-1})
\label{questa}
\end{equation}

\noindent In the discrete theory we assume that the spinor field is a $2^{\nu}$ complex
vector defined at each vertex of the dual Voronoi complex, that is to say
a map that to each vertex $\alpha$ associates the $2^{\nu}$ complex
vector $\psi(\alpha)$.

\noindent Let $|\alpha\beta|$ the distance between the two neighboring circumcenters
in $\alpha$ and $\beta$ \cite{Gab}. 
\noindent We are ready to define the covariant derivative
$(\nabla_{\mu}\psi)(\alpha)$ on a lattice

\begin{equation}
(\nabla_{\mu}\psi)(\alpha)\equiv
{ D_{\alpha\beta}\psi(\beta) - \psi(\alpha)\over |\alpha\beta|}
\label{covariant}\;\;\;\;.
\end{equation}

\noindent So far the discrete version of the the action for the coupling between gravity
and fermionic matter can be written in the form (see also \cite{ren})

\begin{eqnarray}
S^{F} \equiv \sum_{\alpha}\Biggl(\sum_{(\alpha\beta), \beta=1,...,n+1}
{i\over 2|\alpha\beta|}\biggl({\overline {\psi(\alpha)}}b^{a}_{\alpha\beta}\gamma_{a}
D_{\alpha\beta}\psi(\beta)\nonumber \\
- \Big({\overline{D_{\alpha\beta}{\psi}(\beta)}}\Big)b^{a}_{\alpha\beta}\gamma_{a}\psi(\alpha)\biggr)-
m V(\alpha){\overline{\psi(\alpha)}}\psi(\alpha)\Biggr)\;\;\;\;.
\label{fermion}
\end{eqnarray}

\noindent Then the quantum measure, which also includes fermionic matter, can be written
as

\begin{eqnarray}
Z=\int e^{-(S+S^{F})}
\prod_{\alpha}\mu(\psi(\alpha))\mu({\overline
{\psi(\alpha)}})\delta(\sum_{\beta=1}^{n+1}b_{\alpha\beta}(\alpha))\nonumber\\
\times\prod_{\alpha\beta}
\delta(b_{\alpha\beta}(\alpha)-D_{\alpha\beta}b_{\beta\alpha(\beta)})
\mu(D_{\alpha\beta})\mu(b_{\alpha\beta})
\label{minestra}
\end{eqnarray}

\noindent where $S$ is the action for pure gravity (\ref{caction}),
$\mu(D_{\alpha\beta})$ the Haar measure on the two-fold covering group
of $SO(n)$, while $\mu(\psi(\alpha))=d\psi(\alpha)$ and 
$\mu({\overline{\psi(\alpha)}})=d{\overline{\psi (\alpha)}}$  are Grassmann variables.

Now on let's focus our reasonings in four dimensions. The two-fold covering group
of $SO(4)$ is $Spin(4)=SU(2)\times SU(2)$. If we denote with

\begin{equation}
S^{F}_{\alpha\beta}\equiv  
{i\over 2|\alpha\beta|}\biggl({\overline{\psi(\alpha)}}b^{a}_{\alpha\beta}\gamma_{a}
D_{\alpha\beta}\psi(\beta)\nonumber \\
- \Big(\overline{D_{\alpha\beta}{\psi}(\beta)}\Big)b^{a}_{\alpha\beta}\gamma_{a}\psi(\alpha)\biggr)\;\;\;\;,
\label{pose}
\end{equation}

\noindent and with

\begin{equation}
S^{F}_{\alpha}\equiv -mV(\alpha){\overline{\psi(\alpha)}}\psi(\alpha)\;\;\;\;,
\label{posino}
\end{equation}

\noindent we can expand $e^{S^{F}_{\alpha\beta}}$ in Fourier series on the compact group $Spin(4)$ \cite{vilenkin}

\begin{equation}
e^{-{S^{F}_{\alpha\beta}}}=\sum_{I}\sum_{ij}c_{I}^{ij}{\;}_{\alpha\beta}
(\psi,{\overline{\psi}},b_{\alpha\beta})
D_{\alpha\beta}^{I}{\;}_{ij}
\label{expanso}
\end{equation}  

\noindent where  

\begin{equation}
c_{I}^{ij}{\;}_{\alpha\beta}(\psi,{\overline{\psi}},b_{\alpha\beta})\equiv
c_{I}^{ij}{\;}_{\alpha\beta}\bigl(\psi(\alpha),{\overline{\psi(\alpha)}},\psi(\beta),{\overline{\psi(\beta)}},b_{\alpha\beta}\bigr)
\equiv d_{I} \int_{Spin(4)}e^{-{S^{F}_{\alpha\beta}}}\left({\overline{D_{\alpha\beta}}}\right)^{I}_{ij}\mu(D_{\alpha\beta})\;\;\;\;,
\label{2exp}
\end{equation}

\noindent with $I$ denoting the irreducible representations of $Spin(4)$ and $d_{I}$
the dimension of the irreducible representation. Then we can rewrite equation (\ref{minestra})
in the following way:

\begin{eqnarray}
Z=\prod_{\alpha\beta}\sum_{I}\sum_{ij}c_{I}^{ij}{\;}_{\alpha\beta}(\psi,{\overline{\psi}},b_{\alpha\beta})D_{\alpha\beta}^{I}{\;}_{ij}
\delta(b_{\alpha\beta}(\alpha)-D_{\alpha\beta}b_{\beta\alpha}(\beta))
\mu(D_{\alpha\beta})\mu(b_{\alpha\beta})\nonumber\\
\prod_{h\supset \alpha\beta}}\sum_{J}c_{J}^{h}\Big(b_{\alpha\beta},b_{\beta\alpha},...,
b_{\delta^{h}_{k}\alpha},b_{\alpha\delta^{h}_{k}},\Big)\chi_{J}\left(D^{h}\right)
\prod_{\alpha}e^{-S_{\alpha}}\delta\bigl(\sum_{\beta=1}^{n+1}b_{\alpha\beta}(\alpha)\bigr)
\mu(\psi(\alpha))\mu({\overline{\psi(\alpha)})\;\;\;\;,
\label{fermfin}
\end{eqnarray}

\noindent in which $\chi_{J}\left(D^{h}\right)$, according to equation (\ref{svilup}), is the character of the $J$th-irreducible
representation of the two-fold covering $D^{h}$ of the holonomy matrix $U^{h}$ around the plaquette $h$. Although this partition 
function seems pretty complicated, it allows a straightforward {\it combinatorial} interpretation. If, as it seems very plausible even if not 
proved yet, Local
Regge Calculus is related to Spin Foam Formalism then the presence of two independent representation indices $I$ and $J$, 
in the path integral (\ref{fermfin}), can be interpretated, from a diagrammatical and combinatorial point of view as well,
as we can associated each irreducible representation $J$ with the {\bf interior} of each plaquette $h$ and each irreducible 
representation $I$ with each dual-edge $\alpha\beta$. This corresponds, graphically, to color the interior of each plaquette and
each edge with, in general, {\bf different} colors! This would imply that the corrispondings Spin Networks, as projection along 
the evolution direction \cite{rovelli} of Spin Foam $2$-complex, of 
gravity with fermionic matter would be characterized by the fact that a  
$Spin(4)$ rapresentation $J$ is associated to each edge and, pairwise, an independent, from $J$, 
$Spin(4)$ representation $I$ is associated to each vertex as well.

 \section{Conclusions}

We have introduced a discrete
theory of gravity in its first-order formalism. We have defined an action as the sum over the hinges of the traces of
the holonomy matrices multiplied by the oriented volumes of the hinges. This action can be
written on the dual Voronoi complex of the original simplicial complex. We can express
the action as a function of the connection matrices
and of the vectors $b_{\alpha\beta}$ which are normal to the faces of the simplices and
whose modulo is proportional to the volume of the faces themself.

\noindent The action is very similar to the Wilson action of lattice gauge theory\cite{Wilson}.
Here the $b_{\alpha\beta}$ have the same role as the $n$-bein in the
continuum theory. Moreover the action is locally invariant under the
action of Lorentz group.

\noindent A quantum measure and the relative partition function has been defined. They are locally invariant
under the action of $SO(n)$. The expansion of the exponential factor, in the partition function, shows that there
are same similarities between Local Regge Calculus and Spin Foam Formalism. However many severe difference seems to
show up as well, mainly based on the fact that, due to the presence of
constraints on the dynamical variable which have to be implemented at quantum level, 
the quantum measure of Local Regge Calculus, as in the case of Regge Calculus
\cite{menotti1}, is {\it highly} non local. This generates differences between the partition function of 
Local Regge Calculus and Spin Foam Formalism. We guess that, since the action of 
Local Regge Calculus converges in {\it the continuum limit} to the Einstein-Hilbert theory, whereas
none have proved yet that Spin Foam Formalism converges to Gravity and Spin Foam Formalism is mainly based  
on many different models, a possible future work would consist of showing how we can derive different models 
used in Spin Foam Formalism from Local Regge Calculus doing suitable approximations. This should demonstate that Local Regge Calculus contains
Spin Foam Formalism.
 We have introduced a coupling of fermionic matter with discrete gravity. This coupling seems very natural and 
might be, in case, a way to introduce Fermionic matter and Gauge Fields as well in Spin Foam 
Formalism. As in the {\it pure}-Gravity case, the expansion of the exponential factor shows 
strict analogies with Spin Foam Formalism but, we have seen, the introduction of fermionic matter 
implies some peculiar lebeling of the $2$-complex which could affect, pairwise, the Spin Foam 
Complex and the relative Spin Network as well.

\begin{acknowledgments}
I would like to thank Alessandro D'Adda for the early and determinant
collaboration on first-order Regge Calculus, Hugo A. Morales Tecotl for discussion and Carlo Rovelli for collaboration on 
the last part of this paper. I would like to express my gratitude
to Fr. Bill Stoeger S.J. for directing, encouraging, and advising me.
It is a pleasure to thank Fr. George Coyne S.J. and Fr. Francesco Tata S.J. for contributions
to this research.
\end{acknowledgments}

\end{document}